# THE VALUE OF INFORMATION AND CIRCULAR SETTINGS

STEFAN BEHRINGER

*Department of Economics, Universität Bielefeld and Sciences Po*

ROMAN V. BELAVKIN

*School of Sciences and Technology, Middlesex University London*

We present a universal concept for the *Value of Information (VoI)*, based on the works of Claude Shannon's and Ruslan Stratonovich that can take into account very general preferences of the agents and results in a single number. As such it is convenient for applications and also has desirable properties for decision theory and demand analysis. The Shannon/Stratonovich *VoI* concept is compared to alternatives and applied in examples. In particular we apply the concept to a circular spatial structure well known from many economic models and allow for various economic *transport costs*.

## 1. INTRODUCTION

The Value of Information *(VoI)* measures the usefulness or worth of acquiring or possessing certain information. It helps to make informed decisions by assessing the potential benefits that can be gained from obtaining specific information.

Knowing the *VoI* of different tasks allows to prioritize and allocate resources effectively. It helps to determine whether it is worth investing time, effort, or money in order to gather additional information before making a decision. As such the *VoI* allows to weigh the potential benefits against the costs associated with obtaining or processing that information.

Wherever decisions have to be made in business, finance, and research generally, a *VoI* can help to optimize choices and minimize risks. Decisions based on *VoI* allow for more informed and rational decisions by considering the potential impact and value that new information can provide. Ultimately, the *VoI* empowers to make better decisions, reduce uncertainty, and increase the likelihood of achieving desired outcomes.

Unfortunately the problem of how to value and price information within a single *VoI* concept (and ideally by a single number) is unresolved. Decision makers with different preferences and priors of the state of nature will prevent comparisons of statistical experiments using Blackwell's (1951) theorem. Being based on the idea of *stochastic dominance* the information ranking resulting from the theorem cannot tell us how the value of a piece of information changes with respect to the agent's preferences. It also requires decision problems where the set of actions and so the state space is finite or at least countable.[1]

Because of these challenges transfers of a *VoI* concept into economic theory involving rational agents following the work of Blackwell, (1951)(9) has been sparse despite the obvious

---

Stefan Behringer: stefan@stefanbehringer.com

Roman V. Belavkin: R.Belavkin@mdx.ac.uk

We thank Pierpaolo Battigalli, Tilman Börgers, Roberto Serrano, Philipp Strack, Martino Trassinelli, and participants at MaxEnt2022 at IHP Paris for comments.

[1]Prior and preference restrictions are investigated in e.g., Lehmann (1988)(19), Cabrales et al. (2013)(11) and inexpensive information in Moscarini and Smith (2002)(23). De Lara and Gossner (2020)(13) provide a cardinal value by linking experiments to outcomes and Frankel and Kamenica (2019) (15) an axiomatic ex-post approach.

attractiveness of the concept for decision theory. Most of the attention has focused on the demand side properties of the concept, e.g., Arrow (1971)(2), Marshak (1972)(22), Chade and Schlee (2002)(12), Behringer (2021a,b)(4)(5). More recently the focus has been extended to the cost side of producing information, see Denti (2022)(14), or Pomato et al.(24).

A major concern of Radner and Stiglitz (1984)(25), have been non-concavities of the *VoI* concept, i.e. the possibility of increasing marginal returns over some range which poses problems for existence and uniqueness of Nash equilibria and its demand for information in partial- and general-equilibrium settings. Another concern is the possibility of negative values, even in non-strategic interaction.[2]

In this paper we present a universal *VoI* concept that can accommodate the shortcomings of alternative concepts. In particular it can be applied to any decision problem under uncertainty under any experimental signal structure, allows for arbitrary agent preferences and priors, an infinite set of decisions, and even for strategic interaction. The concept is founded in the works of Claude Shannon and Ruslan Stratonovich and the resulting subject of *Information Theory*, see Shannon (1948)(30), Stratonovich (2020,1965)(33),(32). The Shannon/Stratonovich approach leads to a *VoI* that is always non-negative and always retains its concavity properties.

Being based on *entropy considerations*, Shannon's concept of information features prominently in the economic literature on *Rational Inattention* (see Sims (2003)(31), Matejka and McKay (2015)(20) and surveys in Gabaix (2019)(16) or Maćkowiak et al. (2023)(21)). Most recently the idea of flexible information choice has also been applied to strategic situations, see e.g., Yang (2015)(35), Ravid (2017)(26), Behringer (2023)(6).

In order to show the applicability of the concept to economic settings we extend a simple example from Stratonovich (1965) where, as in the literature on *Rational Inattention*, optimization takes place over Blackwell experiments allowing for a choice of what information to receive. We then derive the Shannon/Stratonovich *VoI* in a generic *circular spatial setting* that is familiar from existing economic theory models.

The circle is a convenient and general vehicle to introduce real or perceived space into economic modelling. This is due to two important geometrical properties, its compactness and its absence of boundary. The circle may be subdivided into different segments with the resulting spacial network structure allowing for the occurrence of a *transport cost*, i.e. in the simplest case of *physical space*, the *taxi cost* that occurs when moving a vehicle between segments. Varying these *transport cost* allows for the adjustment of the real or perceived spatial impact on welfare as adapted to particular economic settings.

For this simple model the *VoI* provides an answer to the question how much a decision maker (the taxi driver), given her preferences (the transport costs and priors), is willing to invest to get information about the statistical properties of the random variable (the location of the passenger) rather than experiment (locate her taxi) at random.

A well established model where the circle is employed to model *perceived space* in economic theory is the Salop circle Salop(1979)(29) in price theory that is frequently used to model imperfect and monopolistic competition.

An application of a circle representing *physical space* appears in environmental and resource economics. In the dynamic model of Behringer and Upmann (2014)(3) natural resources are growing on the periphery of a circle so that harvesting requires a decision about the speed and the magnitude of sustainable harvesting.[3] Applications also comprise macro economics. Extensions of growth-theory to spatial dimensions also employ the circle see e.g. Brock and Xepapadeas (2008)(10) surveyed in Augeraud-Véron et al. (2019)(1).

---

[2] see De Lara and Gossner (2020)(13) or Kops and Pasichnichenko (2023)(18).
[3] These results are generalized in Zelikin et al. (2017)(36).

Within the generic circular economic application, the Shannon/Stratonovich *VoI* is then compared to a simpler uniform information concept (the *Hartley information*) that often appears in game and decision theoretic applications (e.g., Hirshleifer and Riley (1979)(17)), and various alternative specifications of agent preferences and the stochastic environment are investigated and comparative statics and limiting bounds obtained.

## 2. THE VALUE OF INFORMATION

Following the approach of the Stratonovich (1965,2020) (32),(33), we let $\Omega = X \times U$ denote sets of random variables. We optimize a utility function $U(x,u)$ over the joint probability measure $p(u,x)$ where $x = x_i, i = 1....n$ is a random variable and $u = u_j, j = 1....n$ some action. This maximization is subject to an information constraint where Shannon information is found as the difference of entropies, see Shannon (1948)(30), the Kullback-Leibler divergence. The total amount of Shannon information is found by integrating over the joint measure.

We thus have the program:

$$U(I) \equiv \sup \left[ E_{p(u,x)} \{U(x,u)\} \text{ s.t. } I_{xy} \leq I \right]$$

$$= - \inf_{p(u,x)} \left[ \int_{X,U} c(x,u)p(x,u)dxdu \right] \text{ s.t. } \int_{X,U} \ln \left[ \frac{p(x,u)}{p(x)p(u)} \right] p(x,u)dxdu \leq I$$

where we assume the product measure and $p(x)p(u)$ as the reference measure in Shannon information where $p(x)$ and $p(u)$ are the marginals.

We now *marginalize*, the joint measure assuming that the $p(x)$ is its marginal when integrating over the action set $U$, i.e.

$$\int_U p(x,u)du = p(x) \quad \text{(C)}$$

so that the *a-priori* distribution is unchanged. Note that $p(x)$ need not be normalized which then implies that the optimal joint measure $p(x,u)$ is not normalized either. Normalization satisfies the following fundamental Lemma:

LEMMA 2.1: *Let $X, Y$ and $\langle \cdot, \cdot \rangle : X \times Y \to \mathbb{R}$ be the dual pair of spaces with non-degenerate bilinear form. Consider the normalization map $N : Y \setminus \{1\}_\perp \to Y$:*

$$N[y] = \frac{y}{\langle 1, y \rangle}.$$

*The domain of $N$ is the complement $Y \setminus \{1\}_\perp$ of annihilator $\{1\}_\perp = \{y : \langle 1, y \rangle = 0\}$ of the element $1 \in X$, (i.e. $\{y : \langle 1, y \rangle \neq 0\}$. The image of $N$ is an affine space $\{y : \langle 1, y \rangle = 1\}$. Normalization then has the following properties:*

$$\langle 1, N[y] \rangle = 1$$

$$N \circ N = N$$

$$N[c \cdot y] = N[y], \forall c \in \mathbb{R} \setminus \{0\}.$$

We employ a Lagrangian dual approach (c.f. Rockafellar (1989)([27]) p.18ff) with *marginalization constraint* ([C]):

$$\mathcal{K} = \int_{X,U} \ln\left[\frac{p_{ij}(x,u)}{p(x)p(u)}\right] p_{ij}(x,u)dxdu +$$

$$\beta\left(\int_{X,U} c(x,u)p_{ij}(x,u)dxdu - C\right) + \sum_{x,u}\gamma_{ij}\left(\int_U p_{ij}(x,u)du - p(x)\right)$$

with a set of partial derivatives that give a set of necessary conditions (i-iii) for a zero-gradient $\nabla\mathcal{K}(x,u,\beta,\gamma_{ij})$. Maximizing over the joint measure $p_{ij}(x,u)$ we find (i)

$$\frac{\partial\mathcal{K}(p(x,u),\beta)}{\partial p_{ij}(x,u)} = \ln\left[\frac{p_{ij}(x,u)}{p(x)p(u)}\right] + 1_{ij} + \beta c(x,u) + \gamma_i(x)$$
$$= \beta c(x,u) - \ln p(x) - \ln p(u) + \ln p(x,u) + 1_{ij} + \gamma_i(x) = 0_{ij}$$

where $1_{ij}$ and $0_{ij}$ are planes with normal vectors $(0,0,1)$ in $(x,u,z)$-space at distances 1 and 0 from the origin. Also:

$$\frac{\partial\sum_{x,u}\gamma_{ij}\left(\int_U p_{ij}(x,u)du\right)}{\partial p_{ij}} = \frac{\partial\sum_{x,u}\gamma_{ij}\left(p_i(x)\right)}{\partial p_i} = \gamma_i(x)$$

Thus (i) may be written as

$$p(x,u) = e^{-\gamma_i(x)-\beta c(x,u)-1_{ij}} p(x)p(u)$$

and necessary conditions (ii-iii) are then

$$\frac{\partial\mathcal{K}(p(x,u),\beta)}{\partial\beta} = \int_{X,U} c(x,u)p(x,u)dxdu - C = 0 \qquad (ii)$$

and

$$\frac{\partial\mathcal{K}(p(x,u),\beta)}{\partial\gamma_i(x)} = \int_U p(x,u)du - p(x) = 0_i. \qquad (iii)$$

LEMMA 2.2: *For the optimal partially normalized measure according to ([C]), integration over U*

$$\int_U p^{p-norm}(x,u)du = p(x)$$

*implies (and is implied by)*

$$\int_U e^{-\gamma_i(x)-\beta c(x,u)} p(u)du = 1. \qquad (1)$$

*Integration over X*

$$\int_X p^{p-norm}(x,u)dx = p(u)$$

*and implies (and is implied by)*

$$\int_X e^{-\gamma_i(x) - \beta c(x,u)} p(x) dx = 1. \tag{2}$$

PROOF: See Appendix. *Q.E.D.*

We also have:

LEMMA 2.3: *Given two sets X and Y with $\Omega = X \times Y$. Partial normalization of some function f(x,y) w.r.t. y satisfies*

$$N_y [k(x) f(x,y)] = N_y [f(x,y)]$$

*and is thus invariant to multiplication by some function c(x).*

PROOF: See that partial normalization w.r.t. y implies:

$$N_y [k(x) f(x,y)] = \frac{k(x) f(x,y)}{\int_Y k(x) f(x,y) dx} = \frac{f(x,y)}{\int_Y f(x,y) dx} = N_y [f(x,y)].$$

*Q.E.D.*

We further find:

LEMMA 2.4: *The optimal measure, with marginalization is*

$$p^{p-norm}(x,u) = e^{-\gamma_i(x) - \beta c(x,u)} p(u)$$

*where*

$$\gamma_i(x) = \frac{\partial \sum_{x,u} \gamma_{ij} \left( \int_U p_{ij}(x,u) du \right)}{\partial p_{ij}} = -\ln \left[ \int_U e^{-\beta c(x,u)} p(u) du \right]$$

*is the Lagrange multiplier (a function in $z - x$ space) of the marginalization constraint ([C]).*

PROOF: See Appendix. *Q.E.D.*

THEOREM 2.5: *The condition for the optimal joint measure $p^{p-norm}(x,u)$ can be derived as*

$$\Gamma(\beta) - \beta \Gamma'(\beta) = -I_{x,u}$$

*when marginalized according to ([C]) where $\Gamma(\beta) = \ln Z(\beta)$ is the cumulant generating function, $Z(\beta)$ is the partition function (*Zustandssumme*), and $\beta$ the inverse Lagrange multiplier of the information constraint. Also*

$$\Gamma'(\beta) = E_{x,u} \{c(x,u)\}.$$

PROOF: In Lemma 2.4 the optimal marginalized measure was found as

$$p^{p-norm}(x,u) = p(u)e^{\underbrace{-\ln\left[\int_U e^{-\beta c(x,u)}p(u)du\right]}_{\gamma_i(x,\beta)} - \beta c(x,u)}$$

for which

$$\frac{\partial \gamma_i(x,\beta)}{\partial \beta} = \frac{\int_U e^{-\beta c(x,u)}p(u)(-c(x,u))du}{\int_U e^{-\beta c(x,u)}p(u)du}$$

$$= \frac{\int_U e^{-\gamma_i(x)-\beta c(x,u)}p(u)p(x)(-c(x,u))du}{p(x)}$$

$$= -\int_U \frac{p^{p-norm}(x,u)}{p(x)}c(x,u)du.$$

from Lemma 2.2 and Lemma 2.3.

Taking logs of $p^{p-norm}(x,u)$ we have

$$\gamma_i(x,\beta) + \beta c(x,u) = -\ln\left[\frac{p(x,u)}{p(x)p(u)}\right]$$

Multiplying by $p^{p-norm}(x,u)$ and integrating over $X$ and $U$ we have

$$\int_{X\times U} p^{p-norm}(x,u)\gamma_i(x,\beta)dxdu + \beta \int_{X\times U} p^{p-norm}(x,u)c(x,u)dxdu =$$

$$-\int_{X\times U} p^{p-norm}(x,u)\ln\left[\frac{p(x,u)}{p(x)p(u)}\right]dxdu$$

so using Fubini's theorem and (C) again we find

$$\int_X \gamma_i(x,\beta)p(x)dx - \beta \int_X \frac{\partial \gamma_i(x,\beta)}{\partial \beta}p(x)dx = -\int_{X\times U} p^{p-norm}(x,u)\ln\left[\frac{p(x,u)}{p(x)p(u)}\right]dxdu.$$

Denoting the cumulant generating function as $\Gamma(\beta) \equiv \int_X \gamma_i(x,\beta)p(x)dx$ then

$$\Gamma(\beta) - \beta\frac{\partial \Gamma(\beta)}{\partial \beta} = -E_{x,u}\left\{\ln\left[\frac{p(x,u)}{p(x)p(u)}\right]\right\}$$

or

$$\Gamma(\beta) - \beta\frac{\partial \Gamma(\beta)}{\partial \beta} = E_{x,u}\left\{I(x,u)\right\} = -I$$

as $\beta$ is the multiplier in the dual, $\beta^{-1}$ is the multiplier of the information constraint and $\Gamma'(\beta) = E_{x,u}\left\{c(x,u,\cdot)\right\} = C$. *Q.E.D.*

Note that $p(x)$ need not be normalized. The analysis may similarly be performed if instead of joint measures conditional measures are employed that are normalized by definition (see Stratonovich (1965) (32). Because the Hamiltonian is the Legendre-Fenchel transform (see Tikhomirov §2 (2011),(34) of the Lagrangian one may alternatively show from the equation of motion that the $\beta^{-1}$ is the multiplier of the information constraint.

The optimal measure can be written as

$$p^{p-norm}(x,u) = \frac{e^{-\beta c(x,u)}}{\int_U e^{-\beta c(x,u)} p(u) du} p(x) p(u) = \frac{e^{-\beta c(x,u)}}{e^{\gamma(\beta,x)}} p(x) p(u)$$

and if the linear transformation $G[\cdot] = \int_X e^{-\beta c(x,u)} (\cdot) dx$ has an inverse, then from (2) in Lemma 2.2 we have that

$$G\left[\frac{p(x)}{\int_U e^{-\beta c(x,u)} p(u) du}\right] = G\left[\frac{p(x)}{e^{\gamma(\beta,x)}}\right] = 1$$

$$\Leftrightarrow$$

$$G^{-1}[1] = \int_U g^{-1}(x,u) du = \frac{p(x)}{e^{\gamma(\beta,x)}}.$$

This leads to the final lemma for an *independence condition* of $p(x)$ and the normalizer $e^{\gamma(\beta,x)}$ of the optimal measure:

LEMMA 2.6:

$$\frac{p(x)}{e^{\gamma(\beta,x)}} = Z^{-1} \in \mathbb{R}\setminus\{0\} \Leftrightarrow G[1] = [G^{-1}[1]]^{-1} = Z = \frac{e^{\gamma(\beta,x)}}{p(x)} \in \mathbb{R}\setminus\{0\}.$$

*so that* the two *marginalization conditions follow:*

$$\int_X p^{p-norm}(x,u) dx = p(u) \text{ and } \int_U p^{p-norm}(x,u) du = p(x).$$

*Additionally*

$$G[1] = \int_X e^{-\beta c(x,u)} dx = Z \text{ and } G^{-1}[1] = \int_U e^{\beta c(x,u)} du = Z^{-1}.$$

PROOF: See Appendix. *Q.E.D.*

3. THE STRATONOVICH (1965) EXAMPLE

Stratonovich (1965)(32) presents a *discrete* canonical example where there are 8 possible realizations of a random variable $x$ on a circle with circumference 8 and costs, given some choice $u$ of the agent, given linear *transport costs* of the form

$$c(x,u) = \min\{|x-u|, 8-|x-u|\}$$

As the simplest economic application think of the case of a *taxi* driver who wants to pick up a potential passenger in an idealized city or route, i.e. on the periphery of a circle. She is not paid for the pickup and wants to minimize the transport costs as given above that she has for this pickup. She can locate herself optimally and passengers appears at random at equally spaced points with uniform probability. She may also acquire binary information about on which half, quarter, eighth,... of the circle a passenger will appear. What would be her willingness to pay for such information? The answer to this question is the *Hartley VoI* that for 8 and 16 equiprobable passenger locations is represented as dots in Figure 1.

We extend the stochastic environment of this motivating example in to an even number $n$ of uniformly distributed points on the circle which we normalize to the unit circle with circumference $2\pi$.

Stratonovich curtails the *general preferences of the agent* for the uniformly distributed outcomes into :

$$Z(\beta) = \frac{1}{n}\sum_x e^{-\beta c(x,u)} = \frac{1}{n}\left(1 + 2\sum_{x=1}^{\frac{n}{2}-1} e^{-\beta x} + e^{-\beta \frac{n}{2}}\right)$$

$$= \frac{1}{n}\left(1 - e^{-\beta \frac{n}{2}}\right)\coth\left(\frac{\beta}{2}\right).$$

where $\beta$ is the inverse Lagrange multiplier for the information constraint with monotone transformation preserving data of the agents general preferences. We next provide a definition: Let $X$ and $Y$ be linear spaces and let some $g: X \times Y \to \mathbb{R}$, then we call $g$ *translation invariant* if

$$\forall w \in Y \; \exists\, v \in X \text{ such that } g(x+v, y+w) = g(x,y).$$

As the information constraint is an inequality, $\beta^{-1} \geq 0$. The *cumulant generating function* can be derived as

$$\ln Z_n(\beta) = \Gamma_n(\beta).$$

The fact that $\Gamma'_n(\beta) = E_{x,u}\{c(x,u,\cdot)\}$ can be seen by expanding expected costs as

$$\Gamma'_n(\beta) = \frac{1}{Z_n(\beta)} Z'_n(\beta)$$

$$= -\frac{1}{n}\left(\frac{e^{-0\beta}}{\frac{1}{n}\left(1-e^{-\beta\frac{n}{2}}\right)\coth\left(\frac{\beta}{2}\right)}(0) + \frac{\sum_{x=1}^{\frac{n}{2}-1} e^{-\beta x}}{\frac{1}{n}\left(1-e^{-\beta\frac{n}{2}}\right)\coth\left(\frac{\beta}{2}\right)}(2\cdot x) + \frac{e^{-\frac{n}{2}\beta}}{\frac{1}{n}\left(1-e^{-\beta\frac{n}{2}}\right)\coth\left(\frac{\beta}{2}\right)}\left(\frac{n}{2}\right)\right)$$

$$= \frac{n}{2(e^{\beta\frac{n}{2}}-1)} - csch(\beta)$$

with

$$\lim_{n \to \infty} \Gamma'_n(\beta) = -csch(\beta) \text{ if } \beta > 0$$

where $csch(\beta)$ is the hyperbolic cosecant of $\beta$. Alternatively if follows from the Legendre-Fenchel transform. Information is then calculated from

$$I = \beta \Gamma'_n(\beta) - \Gamma_n(\beta).$$

Following Stratonovich (1965)(32) we define the Shannon/Stratonovich *VoI* function as

$$V(I) = U(I) - U(I_{min})$$

where we may set the constant $I_{min}$ to $I(0)$ to satisfy a participation constraint.

LEMMA 3.1: *The cost function of the Stratonovich (1965) Example is translation invariant.*

PROOF: See Appendix. *Q.E.D.*

Plotting $V(I)$ parametrically for $\beta$ yields a *VoI* for the $n$ values of 8 (blue) and 16 (yellow) and the corresponding *Hartley VoI* (black dots) where, to follow Stratonovich (1965)(32) exactly, we use $\log_2$. Also for practical purposes it is important to note that the difference between the two *VoI* concepts will be vanishing asymptotically with the number of points $n$, as shown in Stratonovich(2020)(33).

With a circle that grows with $n$ the *VoI* is found to be *increasing in* $n$, i.e. generates a higher *VoI* for any given level of information $I$. As an investigation of the effect of having more random variables it is more interesting for bounded moments of the underlying cost function in the next section we will look at a unit circle.

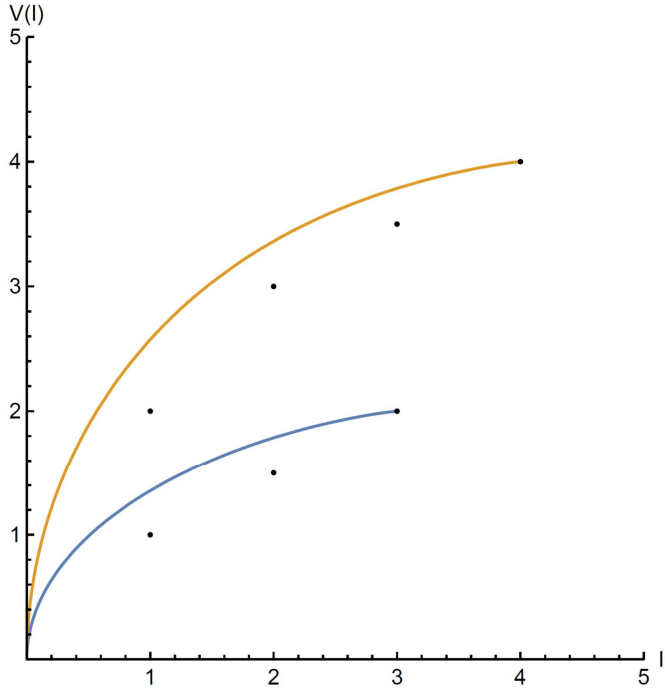

Figure 1.—VoI for circle

## 4. THE EXTENDED STRATONOVICH (1965) EXAMPLE ON A UNIT CIRCLE

Next we look at what happens to the extended Stratonovich example on a *unit* circle, i.e. one where the circumference remains $2\pi$ independently of the number of points $n$ thus limiting the total transport cost effect. The *partition function* for linear cost on the unit circle given some reference measure $p(x)$ under *translation invariance* is given by

$$Z_n(\beta) = \sum_x e^{-\beta c(u-x,0)} p(x)$$

$$= p(x_0) + 2\sum_{x=1}^{\frac{n}{2}-1} p(x_x) e^{-\beta x\left(\frac{2\pi}{n}\right)} + e^{-\beta\pi} p(x_{\frac{n}{2}})$$

To investigate large $n$ cases we need to insure that this function is bounded in $n$ as the partition function is used to normalize an otherwise unbounded measure. General existence conditions for such limits are a investigated in (rigorous) statistical mechanics, e.g. Ruelle (1999)(28). Depending on the choice of the *transport cost* function, and the reference measure (priors) and hence agent preferences, this may impose extra bounds $\beta$. Intuitively we can only consider $\beta$ for which the partition sum is finite as otherwise optimal measures are not well defined. For the linear cost unit circle case there is a set of priors $p(x)$ such that the partition function converges.

LEMMA 4.1: *With linear costs there exists a reference measure p(x) such that the partition function of the Stratonovich (1965) example on the unit circle is finite, i.e.* $\exists \beta^{-1} > 0$ *such that*

$$Z_{lc}(\beta) = \sum_x e^{-\beta c(x-u,0)} p(x) < \infty \forall n \in [1, \infty).$$

PROOF: See Appendix. *Q.E.D.*

LEMMA 4.2: *With $Z_{lc}(\beta)$ the marginal conditions of Lemma 2.2 hold.*

PROOF: See Appendix. *Q.E.D.*

LEMMA 4.3: *The limiting partition function for linear cost and a uniform prior has a closed form solution*

$$\lim_{n\to\infty} Z_{lc}(\beta) = \frac{1-e^{-\beta\pi}}{\beta\pi}.$$

*as does the cumulant generating function* $\Gamma(\beta) = \ln Z(\beta)$.

PROOF: See Appendix. *Q.E.D.*

Figure 2 shows the *VoI* for the $n$ values of 8 (blue) and 16 (yellow), and limit case (green) for the uniform prior. In contrast to the results for the non-normalized circle of section 4, here the *VoI* now decreases for any given value of information. By general properties of the analysis we have:

LEMMA 4.4: *The graph of the Shannon/Stratonovich VoI changes in $n$ according to a cost effect:*

$$\frac{\partial E\{c(\cdot,n)\}}{\partial n} = \frac{\partial^2 \Gamma(\beta,n)}{\partial \beta \partial n}$$

*and a variance effect:*

$$\frac{\partial (\Delta c(\cdot,n))^2}{\partial n} = \frac{\partial^3 \Gamma(\beta,n)}{\partial \beta^2 \partial n}.$$

*We also define $\frac{\partial \Gamma(\beta \to 0, n)}{\partial \beta} \equiv M\{c(x,u,n)\}$ as the Maximum Entropy costs.*

PROOF: By properties of $\Gamma(\beta,n)$, e.g. Theorem 3.5 in Stratonovich(2020)(33). Q.E.D.

*Maximum Entropy* costs $M\{c(x,u,n)\}$ remain constant at $\pi/2$. So there is no impact from $n$ on the *VoI* from changing cost expectations absent information. The total *VoI* to be decreasing in $n$ is driven by the variance effect which is negative, i.e. decreases in $n$. However, as the variance effect converges to zero we find a *lower bound* for the Shannon/Stratonovich *VoI*. Next we will look at cases where *Maximum Entropy* costs vary.

### 4.1. *The extended Stratonovich (1965) Example on the unit circle with non-linear costs*

In contrast to the above we now assume non-linear transport costs. Note that the cost functions remain *translation invariant*, independently of the exact *transport cost function* employed. Non-linearity would be an appropriate choice to model for example *scale economies* in perceived or actual distances on the circle.

We will first investigate the case where these costs are the logarithm of distances. We then have

$$c_{ln}(x,u) = \min\{\ln(|x-u|), \ln(n-|x-u|)\}.$$

The partition function with a uniform prior becomes:

$$Z_{\ln}(\beta) = \frac{1}{n} + \frac{2}{n}\sum_{i=1}^{\frac{n}{2}-1} e^{-\beta \ln\left(i\frac{2\pi}{n}\right)} + \frac{1}{n} e^{-\beta \ln(\pi)}$$

$$= \frac{1}{n} + \frac{2}{n}\sum_{i=1}^{\frac{n}{2}-1} \left(i\frac{2\pi}{n}\right)^{-\beta} + \frac{1}{n}\pi^{-\beta}$$

or

$$Z_{\ln}(\beta) = \left(\frac{2}{n}\right)^{1-\beta} \pi^{-\beta} H\left(\frac{n}{2}-1, \beta\right) + \frac{1+\pi^{-\beta}}{n}$$

with $H(n/2-1,\beta) = \sum_{i=1}^{\frac{n}{2}-1} \frac{1}{i^\beta}$ being the Harmonic number of order $\beta$ which again allows for a closed-form solution. Figure 3 shows the *VoI* for the $n$ values of 8 (blue) and 16 (yellow), and 1024 (green). The *VoI* is increasing in $n$.

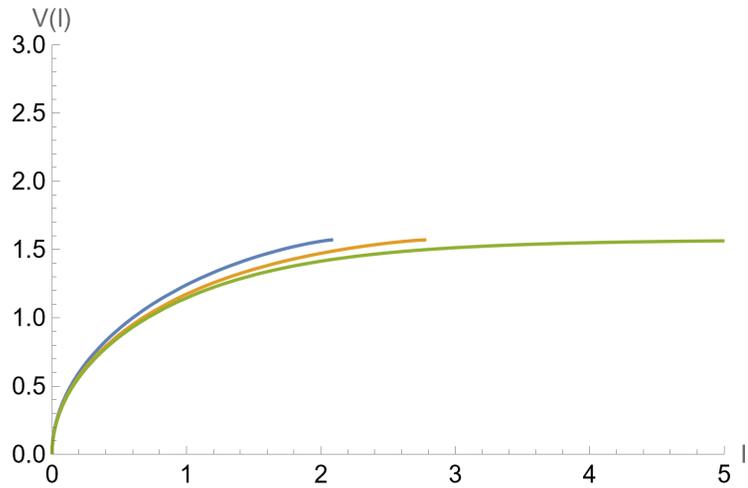

FIGURE 2.—VoI for unit circle

*Maximum Entropy* transport costs of having $n$ points on a unit circle and logarithmic costs between any point and some chosen action are

$$M\{c_{ln}(x,u,n)\} = 0 \cdot \frac{1}{n} + 2 \cdot \frac{1}{n} \sum_{i=1}^{\frac{n}{2}-1} \ln\left(i\frac{2\pi}{n}\right) + \frac{1}{n} \ln \pi$$

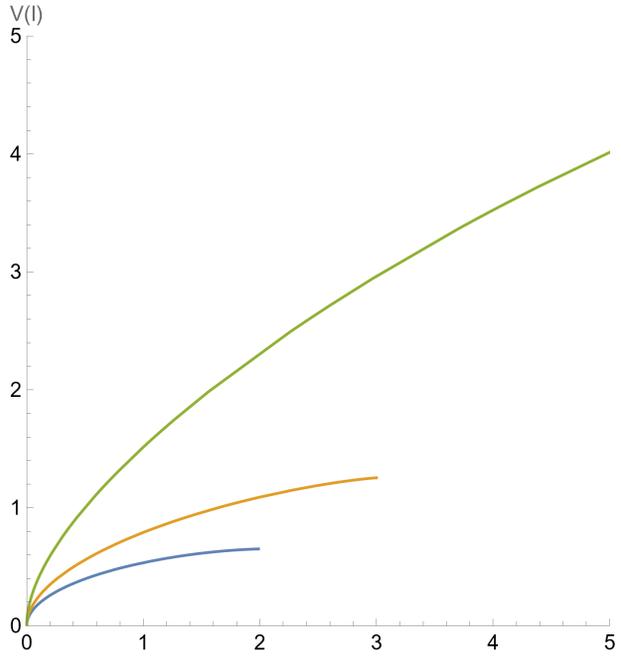

FIGURE 3.—VoI for unit circle with log costs

$$= \frac{1}{n}\left(2\ln\left(\left(\frac{2\pi}{n}\right)^{\frac{n}{2}-1} G\left(\frac{n}{2}\right)\right) + \ln(\pi)\right)$$

where $G$ is the gamma function that is now *decreasing* in $n$. Still as the *VoI* graph is adjusted for the Maximum Entropy costs, (as the *VoI* needs to be zero for any Maximum Entropy costs absent information) there is no direct effect on the graph at the origin. Still the cost effect impacts the variance effect for positive amounts of information that in this case is *increasing in* $n$, for n>5, so the *VoI* graph is increasing.

Again Maximum Entropy costs converge as for a large number of points a limit can be found as
$$\lim_{n \to \infty} M\{c_{ln}(x,u,n)\} = \ln \pi - 1.$$

Similar concave transformations such as root transport costs of the form
$$c_{root}(x,u) = \min\left\{\sqrt{|x-u|}, \sqrt{n-|x-u|}\right\}$$
yield Maximum Entropy costs
$$M\{c_{root}(x,u,n)\} = 0 \cdot \frac{1}{n} + 2 \cdot \frac{1}{n} \sum_{i=1}^{\frac{n}{2}-1} \sqrt{i\left(\frac{2\pi}{n}\right)} + \frac{1}{n}\sqrt{\pi} =$$
$$\sqrt{\pi}\left(\frac{1}{n} + 2\sqrt{2}\left(\frac{1}{n}\right)^{\frac{3}{2}} H(\frac{n}{2}-1, -\frac{1}{2})\right)$$

*increasing* in $n$ but do not allow for closed form solutions for $Z(\beta)$. Still parametric plots yield information about the exact *VoI* numbers in this setting which can be seen in Figure 4 showing the *VoI* for the $n$ values of 8 (blue) and 16 (yellow), and the 1024 (green) with root costs. Again the total variance effect being negative implies that the *VoI* is now decreasing in $n$.

As mentioned above, finding closed form solutions for $Z(\beta)$ is not generic and depends mutually on the choice of the cost function and the spatial structure and reference measures all of which represent the agent's general preferences over the stochastic environment.

All of the examples investigated in this paper make the assumption of reference measures representing ex-ante indifference (uniform priors). Belavkin (2012)(7) building on an example in Stratonovich (2020)(33) has shown that systems with translation invariant quadratic costs and Normal priors result in closed form solutions for $Z(\beta)$, optimal conditional probabilities that are Gaussian, and a *VoI* in closed form. More general agent preferences and priors leading to closed form *VoI* that allows straightforward practical applications to data thus promise an interesting avenue for future research.

### 4.2. *An example on the real line*

To extend the analysis beyond the circular space the following example assumes a segment of the real line of length $2\pi$ keeping translation invariance so that in the taxi interpretation, we face a *one-way* unit circle. Again all $n$ states of the world have equal probability and the corresponding costs are again linear. The respective partition function is
$$Z_{lc}(\beta) = \frac{1}{n}\left(e^{-\beta\pi} \sinh(\pi\beta)\left(\coth\left(\frac{\pi\beta}{n}\right) + 1\right)\right)$$
and from the cumulant generating function $\Gamma_{ln}(\beta)$ we find
$$U_{ln}(\beta) = \frac{1}{n}\left(\pi\left(n(\coth(\pi\beta)-1) - \coth\left(\frac{\pi\beta}{n}\right) + 1\right)\right).$$

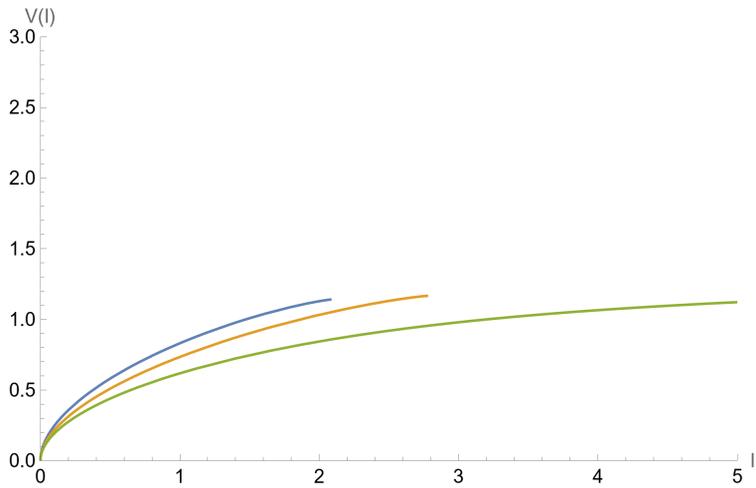

FIGURE 4.—VoI for unit circle with root costs

Information follows again from Theorem 2.5.

Plotting parametrically we obtain *VoI* Figure 5 where the blue graph is the $n=8$ state and the yellow graph is the $n=16$, the green graph the $n=1024$ state case. Again we normalize by the *Maximum Entropy costs* that are increasing in $n$ but bounded on the $2\pi$ line. With a positive variance effect the *VoI* graph is now increasing in $n$ for any given information level even under linear cost in contrast to the *two-way* unit circle.

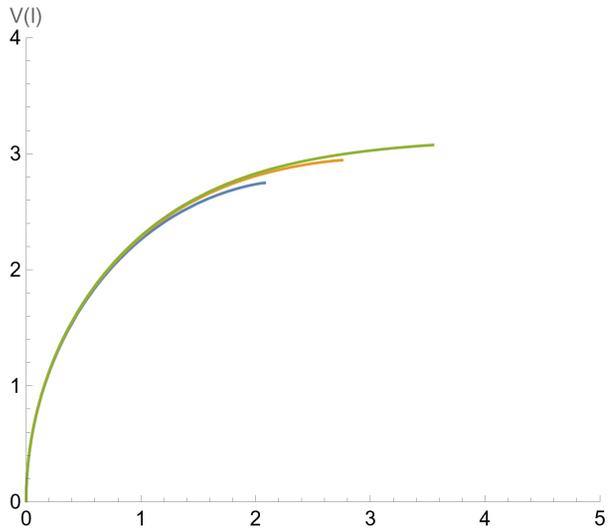

FIGURE 5.—VoI for one-way circle example

## 5. CONCLUSION

We have presented a *VoI* concept that we call the Shannon/Stratonovich *VoI* that overcomes many of the shortcomings of alternative approaches e.g. those based on Blackwell's partial ordering or the one attributed to Hartley. In particular it can be applied to any decision problem under uncertainty under any experimental signal structure, allows for a wide set of general agent preferences and priors, an uncountable set of decisions, and even strategic interaction. The concept avoids issues of non-concavities that are central to economic concerns as put forward prominently in the works of Roy Radner and Joseph Stiglitz (1984)(25).

We have applied the concept to generic economic situations that are set on the periphery of a circle as used for example in industrial organization, spatial macroeconomics, or renewable resource economics and beyond.

We observe that when extending the original example of Stratonovich (1965)(32) to a *unit* circle, i.e. one that is bounded by the same interval of values for varying numbers of options $n$, having more options (e.g. more firm locations, more product varieties, more resource patches) changes the *VoI*. In the absence of information we can characterize Maximum Entropy payoffs that will depend on the spatial structure and the form of the utility/cost functions and priors and so the agent's general preferences. These payoffs also vary with the number of options $n$ for strictly positive information levels and we categorize the underlying effects. The existence of a limit for these effects and hence the *VoI* in $n$ in bounded spaces implies that even if the random variables (locations, varieties, or patches) are best modelled with a density, the *VoI* graph retains a positive slope. Most importantly the above procedure shows how to obtain the unique *VoI* number that allows for comparisons across widely differing settings enabling the decision maker to decide whether or not to acquire additional information.

## APPENDIX

PROOF: OF LEMMA 2.2: Given

$$p(x,u) = e^{-\gamma_i(x) - \beta c(x,u) - 1_{ij}} p(x) p(u)$$

then even if only partially normalized the optimal measure satisfies

$$p^{p-norm}(x,u) = \frac{p(x,u)}{\int_U p(x,u) du} = \frac{e^{-\beta c(x,u) - 1_{ij}}}{\int_U e^{-\beta c(x,u) - 1_{ij}} p(x) p(u)} p(x) p(u) du = \frac{e^{-\beta c(x,u)}}{\int_U e^{-\beta c(x,u)} p(u) du} p(u)$$

by Lemma 2.1 so we may disregard the $(x,u,1)$ plane as normalization pertains in dual vector space. Using the form

$$p^{p-norm}(x,u) = e^{-\gamma_i(x) - \beta c(x,u)} p(x) p(u)$$

and integrating the optimal partially normalized measure over $U$ given the marginalization condition (C) yields

$$\int_U p^{p-norm}(x,u) du = p(x) \int_U e^{-\gamma_i(x) - \beta c(x,u)} p(u) du = p(x)$$

iff

$$\int_U e^{-\gamma_i(x) - \beta c(x,u)} p(u) du = 1$$

Integrating the optimal partially normalized measure over $X$ yields

$$\int_X p^{p-norm}(x,u)dx = p(u)\int_X e^{-\gamma_i(x)-\beta c(x,u)}p(x)dx = p(u)$$

iff

$$\int_X e^{-\gamma_i(x)-\beta c(x,u)}p(x)dx = 1.$$

<div align="right">Q.E.D.</div>

PROOF: OF LEMMA 2.4: By condition i) and ii) of the Lagrangian we can write the optimal measure as

$$p^{p-norm}(x,u) = \frac{e^{-\beta c(x,u)}p(x)p(u)}{\int_U e^{-\beta c(x,u)}p(x)p(u)du} = \frac{e^{-\beta c(x,u)}p(u)}{\int_U e^{-\beta c(x,u)}p(u)du}$$

following Lemma 2.3 and thus

$$p^{p-norm}(x,u) = e^{-\ln\left[\int_U e^{-\beta c(x,u)}p(u)du\right] - \beta c(x,u)}p(u)$$

so that by the Lagrange conditions for the marginalized optimal solution it follows that

$$\gamma_i(x) = -\ln\left[\int_U e^{-\beta c(x,u)}p(u)du\right].$$

<div align="right">Q.E.D.</div>

PROOF: OF LEMMA 2.6: If $\frac{p(x)}{e^{\gamma(\beta,x)}} = Z^{-1} \in \mathbb{R}\setminus\{0\}$, then

$$G\left[\frac{p(x)}{e^{\gamma(\beta,x)}}\right] = G[Z^{-1}] = \int_X e^{-\beta c(x,u)}(Z^{-1})dx = Z^{-1}\int_X e^{-\beta c(x,u)}dx = Z^{-1}G[1] = 1$$

where the last equality follows from by (2) in Lemma 2.2. Note that

$$Z = \int_X e^{-\beta c(x,u)}dx. \tag{I}$$

This gives

$$G[1] = Z = \frac{e^{\gamma(\beta,x)}}{p(x)} \in \mathbb{R}\setminus\{0\}$$

If $G[1] = Z \in \mathbb{R}\setminus\{0\}$, then $G$ is invertible, and using $G^{-1}[1] = \frac{p(x)}{e^{\gamma(\beta,x)}}$ we have

$$1 = G^{-1}[Z] = \int_U g^{-1}(x,u)Zdu = Z\int_U g^{-1}(x,u)du = ZG^{-1}[1] = Z\frac{p(x)}{e^{\gamma(\beta,x)}}$$

from which it follows that $\frac{p(x)}{e^{\gamma(\beta,x)}} = Z^{-1} \in \mathbb{R}\setminus\{0\}$.

For $\frac{p(x)}{e^{\gamma(\beta,x)}} = Z^{-1} \in \mathbb{R}\setminus\{0\}$ we therefore have

$$p^{p-norm}(x,u) = \frac{e^{-\beta c(x,u)}}{\int_U e^{-\beta c(x,u)} p(u) du} p(x) p(u) = \frac{e^{-\beta c(x,u)}}{e^{\gamma(\beta,x)}} p(x) p(u) = \frac{e^{-\beta c(x,u)}}{Z} p(u)$$

with

$$Z = \int_X e^{-\beta c(x,u)} dx = \frac{\int_U e^{-\beta c(x,u)} p(u) du}{p(x)} \tag{II}$$

so that *both* marginalization conditions now follow:

$$\int_X p^{p-norm}(x,u) dx = \int_X \frac{e^{-\beta c(x,u)}}{Z} p(u) dx = \frac{Z}{Z} p(u) = p(u)$$

$$\int_U p^{p-norm}(x,u) du = \int_U \frac{e^{-\beta c(x,u)}}{Z} p(u) du = \frac{e^{-\beta c(x,u)}}{Z} = \frac{e^{-\beta c(x,u)}}{e^{-\beta c(x,u)}} p(x) = p(x).$$

Additionally

$$G[1] = \int_X e^{-\beta c(x,u)} dx = Z \text{ and } G^{-1}[1] = \int_U e^{\beta c(x,u)} du = Z^{-1}.$$

<div align="right">Q.E.D.</div>

PROOF: OF LEMMA 3.1: Translation invariance for costs $c : X \times U \to \mathbb{R}$ is defined as

$$\forall w \in U \ \exists v \in X \text{ such that } c(x+v, y+w) = c(x,y)$$

In the circle example with uniform density the impact of action is nullified by defining costs symmetrically relative to any position, e.g. $w = -y$ for any $u \in U$. Then there is a $v \in X$ such that $c(x,u) = c(x+v, 0)$. This implies that also the transformation $e^{-\beta c(x,u)}$ is translation invariant and

$$\sum_x e^{-\beta c(x,u)} = \sum_x e^{-\beta c(x-u,0)}.$$

With costs defined in absolute value, the residual costs also satisfy

$$\sum_x e^{-\beta c(|x-u|,0)} = \sum_u e^{-\beta c(|x-u|,0)}.$$

<div align="right">Q.E.D.</div>

PROOF: of LEMMA 4.1 and LEMMA 4.3:

The partition function for linear cost on the unit circle given the reference measure $p(x)$ is given by

$$Z(\beta) = \sum_x e^{-\beta c(u-x,0)} p(x)$$

$$= p(x_0) + 2\sum_{x=1}^{\frac{n}{2}-1} p(x_x) e^{-\beta x\left(\frac{2\pi}{n}\right)} + e^{-\beta\pi} p(x_{\frac{n}{2}})$$

$$= \frac{1}{n} + \frac{2e^{-\beta\pi}\left(e^{\beta\pi} - e^{\frac{2\beta\pi}{n}}\right)}{\left(e^{\frac{2\beta\pi}{n}} - 1\right)n} + \frac{1}{n}e^{-\beta\pi}$$

which is finite even for large $n$ if we take the uniform measure $p(x) = n^{-1}$ as

$$\lim_{n\to\infty} \frac{2e^{-\beta\pi}\left(e^{\beta\pi} - e^{\frac{2\beta\pi}{n}}\right)}{\left(e^{\frac{2\beta\pi}{n}} - 1\right)n} = \lim_{n\to\infty} \frac{1}{n}\left(1 - e^{-\frac{\beta(n-2)\pi}{n}}\right)\coth\left[\frac{\beta\pi}{n} - 1\right] = \frac{1 - e^{-\beta\pi}}{\beta\pi}.$$

So there exists a reference measure such that the partition function is finite for all $n \in [1, \infty)$. ∎

*Q.E.D.*

PROOF: OF LEMMA 4.2:

With translation invariance let $C(x) = \sum_u e^{-\beta c(x-u,0)} p(u)$. This depends on the reference prior $p(u)$ that may be any (normalized) measure. So we may choose $p(u)$ such that $C(x) = \sum_u e^{-\beta c(x-u,0)} p(u) = \sum_u g(x-u,0)p(u) = Z$ for $Z \in \mathbb{R}\backslash\{0\}$. For example on the unit-circle with linear residual cost, total costs of all actions are

$$\sum_u c(|x-u|,0) = \sum_x c(|x-u|,0) = 0 + 2\sum_{x=1}^{\frac{n}{2}-1} x\left(\frac{2\pi}{n}\right) + \pi = \frac{1}{2}\pi(n-2) + \pi = \frac{1}{2}\pi n$$

so that average costs are

$$\sum_x c(x-u,0)p(x) = \pi/2 \text{ if } p(x) = 1/n.$$

independent of $x$. Then

$$\lim_{n\to\infty}\sum_x g(x-u,0)p(x) = \frac{1-e^{-\beta\pi}}{\beta\pi} < \infty \text{ for some } \beta^{-1} > 0 \text{ and } p(x) = 1/n$$

so also

$$\lim_{n\to\infty}\sum_u g(x-u,0)p(u) = Z_\infty$$

is bounded for reference measure $p(u) = 1/n$ and some $\beta^{-1} > 0$. The optimal joint measure was found as

$$p(x,u) = \frac{p(x)g(x,u)p(u)}{C(x)} = \frac{p(x)g(x,u)p(u)}{Z}$$

summing over all $x$ yields

$$\sum_x \frac{p(x)g(x,u)}{Z} = 1$$

due to partial marginalization w.r.t. $x$. Then assuming the transformation $L[.] = \sum_x g(x,u)[.]$ has an inverse $L^{-1}[.]$ we have

$$L\left[\frac{p(x)}{Z}\right] = Z^{-1}\sum_x g(x-u,0)p(x) = 1 \text{ iff } L^{-1}[1] = \frac{p(x)}{Z}$$

and thus with linear residual costs the partition function is:

$$Z(\beta) = \sum_x g(x-u,0)p(x)$$

which is bounded for all $n$ for some $\beta^{-1} > 0$. Also

$$\sum_x p(x,u) = \sum_x \frac{p(x)g(x,u)p(u)}{Z} = \frac{Zp(u)}{Z} = p(u)$$

and

$$\sum_u p(x,u) = \sum_u \frac{p(x)g(x,u)p(u)}{Z} = \frac{p(x)C(x)}{Z} = p(x).$$

*Q.E.D.*

## REFERENCES

[1] AUGERAUD-VERON, EMMANUELLE, RAOUF BOUCEKKINE, AND VLADIMIR VELIOV (2019): "Distributed Optimal Control Models in Environmental Economics: A Review", *Math. Model. Nat. Phenom.*, **14**, 106, 1-14. [2]

[2] ARROW, KENNETH (1971): "The Value of and Demand for Information", in *Decision and Organization*, ed. by C.B. McGuire and R. Radner. Amsterdam: North-Holland. [2]

[3] BEHRINGER, STEFAN, AND THORSTEN UPMANN (2014): "Optimal harvesting of a spatial renewable resource", *Journal of Economic Dynamics and Control*, **42**, 105-120. [2]

[4] BEHRINGER, STEFAN (2021a): "Multiplicative Normal Noise and Nonconcavity in the Value of Information", *Theoretical Economics Letters*, **11**, 116-124. [2]

[5] BEHRINGER, STEFAN (2021b): "Expanding Multi-Market Monopoly and Nonconcavity in the Value of Information ", arXiv. https://arxiv.org/abs/2111.00839 [2]

[6] BEHRINGER, STEFAN (2023): "Value of Information in Zero-Sum games ", mimeo, submitted. [2]

[7] BELAVKIN, ROMAN (2012): "Optimal measures and Markov transition kernels", in *Journal of Global Optimizaiton*, **55**, 387-416. [15]

[8] BELAVKIN, ROMAN (2022): "Value of Shannon's Information Examples and On Normalization of Measures ", mimeo, University of Nottingham. []

[9] BLACKWELL, DAVID (1951): "The Comparison of Experiments", in Proceedings of the Second Berkeley Symposium on Mathematical Statistics and Probability, ed. by J. Neyman, University of California Press, Berkeley, 93–102. [1]

[10] BROCK, WILLIAM, AND ANASTASIOS XEPAPADEAS, (2008): "Diffusion-induced instability and pattern formation in infinite horizon recursive optimal control", *Journal of Economic Dynamics and Control*, **32**, 2745-2787. [2]

[11] CABRALES, ANTONIO, OLIVIER GOSSNER, AND ROBERTO SERRANO (2013): "Entropy and the Value of Information for Investors", *American Economic Review*, **102**, 360-377. [1]

[12] CHADE, HECTOR, AND EDWARD SCHLEE (2002): "Another look at the Radner-Stiglitz nonconcavity in the value of information", *Journal of Economic Theory*, **107**, 421-452. [2]

[13] DE LARA, MICHELE, AND OLIVIER GOSSNER (2020): "Payoff-Beliefs Duality and the Value of Information", *SIAM Journal on Optimization*, **30**, 1, 464-489. [1, 2]

[14] DENTI, TOMMASO (2022): "Posterior Separable Cost of Information ", *American Economic Review*, **112**, 3215-3259. [2]

[15] FRANKEL, ALEXANDER, AND EMIR KAMENICA (2019): "Quantifying information and uncertainty", *American Economic Review*, **109**, 3650-3680. [1]

[16] GABAIX, XAVIER (2019): "Behavioral Inattention", in *Handbook of Behavioural Economics: Applications and Foundations*, ed. by. Bernheim B.D., DellaVigna, S., and Laibson, D., Elsevier, 2019, Volume 2, Chapter 4, 261-343. [2]

[17] HIRSHLEIFER, JACK, AND JOHN RILEY (1979): "The Analytics of Uncertainty and Information - An Expository Survey", *Journal of Economic Literature*, $XVII$, 1375-1421. [3]

[18] KOPS, CHRISTOPHER, and ILLIA PASICHNICHENKO (2023): "Testing negative value of information and ambiguity aversion", *Journal of Economic Theory*, **213**, 1-31. [2]

[19] LEHMANN, ERICH (1988): "Comparing location experiments", *Annals of Statistics*, **16**, 521-533. [1]

[20] MATEJKA, FILIP AND ALISDAIR MCKAY, (2015): "Rational Inattention in Discrete Choices: A New Foundation for the Multinomial Logit Model", *American Economic Review*, **105**, 272-298. [2]

[21] MACKOWIAK, BARTOSZ, FILIP MATEJA, AND MIRKO WIEDERHOLT (2023): "Rational Inattention: A Review", *Journal of Economic Literature*, **61**, 226-73. [2]

[22] MARSHAK, JACOB, AND ROY RADNER (1972): *Economic Theory of Teams*. Yale University Press. [2]

[23] MOSCARINI, GIUSEPPE, AND LONES SMITH (2002): "The law of large demand for information", *Econometrica*, **70**, 2351-2366. [1]

[24] POMATTO, LUCIANO, PHILIPP STRACK, AND OMER TAMUZ: "The cost of information", in *arXiv*, https://arxiv.org/abs/1812.04211 [2]

[25] RADNER, ROY, AND JOSEPH STIGLITZ (1984): "A nonconcavity in the value of information", in *Bayesian Models of Economic Theory*, ed. by M. Boyer and R. Kilstrom. Elsevier, 1984, 33-52. [2, 18]

[26] RAVID, DORON (2017): "Ultimatum Bargaining with Rational Inattention", *American Economic Review*, **110**, 2948-2963. [2]

[27] ROCKAFELLAR, TERRY (1989): *Conjugate Duality and Optimization, Regional conference series in applied mathematics*, SIAM. [4]

[28] RUELLE, DAVID (1999): *Statistical Mechanics: Rigorous Results*, World Scientific (orig. 1969). [11]

[29] SALOP, STEPHEN (1979). "Monopolistic competition with outside goods",*The Bell Journal of Economics*, **10**, 141-156. [2]

[30] SHANNON, CLAUDE (1948) "A Mathematical Theory of Communication ", *Bell System Technical Journal*, **27**, 379-423. [2, 3]

[31] SIMS, CHRISTOPHER (2003) "Implications of Rational Inattention", *Journal of Monetrary Economics*, **50**, 665-690. [2]

[32] STRATONOVICH, RUSLAN (1965): "On the Value of Information", *Izv. USSR Acad. Sci. Tech. Cybern.*, **5**, 2-12. [2, 3, 7, 8, 9, 18]

[33] STRATONOVICH, RUSLAN (2020): *Theory of Information and its Value* ed. by R.V. Belavkin, P.M. Pardalos, J.C. Principe, original 1975. [2, 3, 9, 12, 15]

[34] TIKHOMIROV, VLADIMIR (2011): "Analysis II", *Convex Analysis and Approximation Theory, Encyclopaedia of Mathematical Sciences*, 14, Theory. ed. by R.V. Gamkrelidze. Springer, original 1990. [7]

[35] YANG, MING (2015): "Coordination with flexible information acquisition", *Journal of Economic Theory*, **158**, 721-738. [2]

[36] ZELIKIN, MICHAIL, LEV LOKUTSIEVSKIY, AND SERGEY SKOPINCEV (2017): "On optimal harvesting of a resource on a circle", *Matematicheskie Zametki*, **102**, 565-578. [2]